\documentclass[pre,aps,onecolumn,superscriptaddress,notitlepage,floatfix,nofootinbib]{revtex4-2}
\usepackage{amssymb}
\usepackage{epsfig}
\usepackage{amsmath}
\usepackage{subfigure}
\usepackage{graphicx}
\usepackage{textcomp}
\usepackage{url}
\usepackage{float}
\usepackage{color}
\usepackage{cases}
\usepackage[normalem]{ulem}

\usepackage[utf8]{inputenc}
\usepackage{cancel}

\usepackage[titletoc,title]{appendix}

\usepackage[colorlinks=true, urlcolor=blue, anchorcolor=blue, citecolor=blue,filecolor=blue,linkcolor=blue,menucolor=blue
]{hyperref}

\usepackage{color}

\newcommand{\be}{\begin{equation}}
\newcommand{\ee}{\end{equation}}
\newcommand{\bea}{\begin{eqnarray}}
\newcommand{\eea}{\end{eqnarray}}

\begin{document}

\title{Fractional Brownian Motion with Negative Hurst Exponent}

\author{Baruch Meerson}
\email{meerson@mail.huji.ac.il}
\affiliation{Racah Institute of Physics, Hebrew University of
Jerusalem, Jerusalem 91904, Israel}

\author{Pavel V. Sasorov}
\email{pavel.sasorov@gmail.com}
\affiliation{ELI Beamlines Facility, Extreme Light Infrastructure ERIC, 252 41 Dolni Brezany,  Czech Republic}

\begin{abstract}
Fractional Brownian motion (fBm) is an important scale-invariant Gaussian non-Markovian process with stationary increments, which serves as
a prototypical example of a system with long-range temporal correlations and anomalous diffusion. The fBm is traditionally defined for the Hurst exponent $H$ in the range \( 0 < H < 1 \). Here we extend this definition to the  regime \( -1/2 < H < 0 \).
The extended fBm  is not a pointwise process, so we regularize it via a local temporal averaging with a narrow filter. The resulting process is both very rough and persistent, that is long-range positively correlated. In addition, this process is stationary. The stationarity implies that diffusion is completely suppressed in this region of $H$.   We also study another closely related Gaussian process: the stationary fractional Ornstein--Uhlenbeck (fOU) process, extended to the range \( -1/2 < H < 0 \) and smoothed in the same way as the fBm. 
Remarkably, the smoothed fOU process is asymptotically insensitive to the strength of the confining potential. Finally, we determine the optimal paths of the  
fBm and fOU processes for \( -1/2 < H < 0 \), conditioned on reaching a specified value when starting from zero.  In the marginal case \( H = 0 \), our results match continuously with known results for the traditionally defined fBm and fOU processes.

\end{abstract}

\maketitle
\tableofcontents

\section{Introduction}

The Mandelbrot -– van Ness fractional Brownian motion (fBm) \cite{Kolmogorov,Mandelbrot,Qian2003} provides a prototypical
example of a random process with long-range temporal correlations, and it has found many applications,
ranging from anomalous diffusion in the living cell to hydrology, telecommunications, and  financial time series \cite{Chow,Metzlerreview}. 
The fBm $x(t)$ is a scale-invariant Gaussian process with stationary increments, which can be defined, for all times $|t|<\infty$,
by its autocorrelation function \cite{Mandelbrot}
\begin{equation}\label{kappafBm}
\kappa(t_1,t_2)\equiv \langle x(t_1) x(t_2)\rangle = D \left(|t_1|^{2H}+|t_2|^{2H}-|t_1-t_2|^{2H}\right)\,,
\end{equation}
where $\langle \dots \rangle$ denotes ensemble averaging.  

The Hurst exponent $H$, which appears in Eq.~(\ref{kappafBm}),  is in the range $0<H<1$.  For $0<H<1/2$ the process is anti-persistent and describes a sub-diffusion, whereas for $1/2<H<1$ it is persistent and describes a super-diffusion.  Our objective here is to extend the fBm process to the range $-1/2<H<0$. It is known that, when the  Hurst exponent of a  scale-invariant process is negative, the process is very singular at short times, see e.g. Ref. \cite{Monthus2026}. On the other hand, as we will see below, the extended fBm, when properly regularized,  has very interesting and unexpected properties, which are worth a detailed analysis and can find useful applications. 

We will regularize the fBm, extended to the regime of $-1/2<H<0$, via a local temporal averaging with a narrow filter, arriving at a process that is both rough and persistent. In addition, this process turns out to be stationary. As a result, diffusion -- the hallmark of the standard fBm for $0<H<1$ -- is completely suppressed in this region of $H$.  

We also extend  to the range \( -1/2 < H < 0 \) the closely related  stationary fractional Ornstein--Uhlenbeck (fOU) process, and regularize it at small scales in the same way as the fBm. We uncover a remarkable phenomenon of asymptotic insensitivity of the  smoothed extended fOU process to the strength of the confining quadratic potential. Finally, we determine the optimal paths of the conditioned fBm and fOU for \( -1/2 < H < 0 \), which provide a useful insight into the physics of these two extended processes.

We start Sec. \ref{main} with a formal definition of the ``bare" (that is, non-regularized) fBm process with  $-1/2<H<0$, which is based
on a straightforward extension of the fractional Gaussian noise (fGn) to this nontraditional region of  $H$. Then we introduce, in Sec. \ref{smoothfBm}, a regularized, or smoothed fBm process and study its properties. In Sec. \ref{smoothfGn} we use the smoothed fBm to reconstruct the smoothed fGn which drives it.  In Sec. \ref{OFM} we study the optimal
(that is, most likely) path of the smoothed fBm process, conditioned on reaching a specified value, when starting from zero. The optimal path dominates the tails of the probability distribution of the conditioned process and provides an additional insight into the problem. Section \ref{fOU} addresses the fractional Ornstein-Uhlenbeck
(fOU) process, extended to the regions of $-1/2<H<0$ in a similar manner. Finally, Sec. \ref{discussion} presents a brief discussion of our results.

\section{Fractional Gaussian noise and fractional Brownian motion  for  $-1/2<H<0$}
\label{main}

\subsection{Bare process}
\label{bare}

The fBm $x(t)$ in one dimension can be described in terms of the Langevin equation
\begin{equation}\label{Langevin0}
\dot{x} = \xi(t)\,,
\end{equation}
where $\xi(t)$ is the centered fractional Gaussian noise (fGn) \cite{Mandelbrot}. In the traditional range $0<H<1$, the autocorrelation function of the fGn,  
$c(t_1-t_2)=\langle \xi(t_1) \xi(t_2)\rangle$ is well known \cite{Mandelbrot,Qian2003}:
\begin{equation}
c(\tau)
 = 2DH\frac{d}{d\tau}\left(|\tau|^{2H-1}\text{sign}\,\tau\right)
 \equiv D \frac{d^2}{d\tau^2}|\tau|^{2H}\,,\quad 0<H<1\,.
 \label{corr}
\end{equation}
The spectral density of the fGn -- the Fourier transform $c_{\omega}$ of $c(\tau)$ is 
\begin{equation}
\label{comega}
c_{\omega} = \int_{-\infty}^{\infty}dt\, c(t) e^{i\omega t}=2 D\sin (\pi  H)\Gamma (2 H+1)|\omega| ^{1-2H}  \,,\quad 0<H<1\,,
\end{equation}
where $\Gamma(z)$ is the gamma function.

One can notice that the right hand side in Eq.~(\ref{comega}) remains well defined also for $-1/2<H<0$, except that the spectral density is negative here because of the factor $\sin (\pi  H)$.  We formally define the fGn in the range $-1/2<H<0$ by simply changing this sign and postulating the following autocorrelation function:
\begin{equation}
c(\tau)
 = - 2DH\frac{d}{d\tau}\left(|\tau|^{2H-1}\text{sign}\,\tau\right)
 \equiv - D \frac{d^2}{d\tau^2}|\tau|^{2H}\,,\quad -1/2<H<0\,,
 \label{corrless}
\end{equation}
or the spectral density
\begin{equation}
\label{comegaless}
c_{\omega} = \int_{-\infty}^{\infty}dt\, c(t) e^{i\omega t}=2 D|\sin (\pi H)|\Gamma (2 H+1)|\omega| ^{1-2H}  \,,\quad -1/2<H<0\,.
\end{equation}

Now we can use the Langevin equation~(\ref{Langevin0}) to extend  the definition of the fBm to the non-traditional range $-1/2<H<0$.  By virtue of its Gaussianity, the (centered) extended fBm is fully determined by its autocorrelation function  $\kappa(t_1,t_2)\equiv \langle x(t_1) x(t_2)\rangle$,  which we will now calculate.  Using Eqs.~(\ref{Langevin0}) and (\ref{corrless}) and formally demanding that $x(t=-\infty) = 0$,  we obtain
\begin{eqnarray}\label{kappa0}
\kappa(t_1,t_2)&=&\int_{-\infty}^{t_1} d\tau_1 \int _{-\infty}^{t_2} d\tau_2 \,\langle \xi(\tau_1) \xi(\tau_2) \rangle\ = \int_{-\infty}^{t_1} d\tau_1 \int _{-\infty}^{t_2} d\tau_2 \,c(\tau_1-\tau_2)  \nonumber \\
&=& D \int_{-\infty}^{t_1} d\tau_1 \int _{-\infty}^{t_2} d\tau_2 \,\frac{\partial^2}{\partial\tau_1\partial\tau_2} |\tau_1-\tau_2|^{2H}= D |t_1-t_2|^{2H}\,,\quad -1/2<H<0\,.
\end{eqnarray}
Note that  at $H<0$, the boundary terms at $-\infty$, resulting from the integrations over $\tau_1$ and $\tau_2$, vanish. 

The autocorrelation function (\ref{kappa0}) describes a long-range positively correlated -- hence persistent-- scale-invariant process. In contrast to the standard fBm -- a nonstationary process with stationary increments -- the extended fBm is a \emph{stationary} process. It is, however, excessively rough: it has an infinite variance of fluctuations in every point, so it is not a point-wise process \cite{Monthus2026}. Here we note that the nonexistence of well-behaved  ``bare" one-point statistics (sometimes referred to as an ultraviolet catastrophe) is often encountered in macroscopic models of stochastic processes and fields, such as fluctuating growing interfaces \cite{Krug,HalpinHealy2014,Almeida,Reis,Carrasco,SmithMeersonSasorov}, mass or energy transport in diffusive lattice gases \cite{BM2024b}, and other systems.

We will use a standard remedy against the ultraviolet catastrophe shortly. Prior to that, however, we present an equivalent alternative definition of the extended fBm.  It employs a Riemann-Liouville fractional integral -- a convolution of the white Gaussian noise $\eta(t)$ [so that $\langle \eta(t_1) \eta(t_2)\rangle = \delta(t_1-t_2)$] with a  power-law kernel:
\begin{equation}\label{altdef}
x(t) = a_H \int_{-\infty}^{t} \left(t-\tau\right)^{H-\frac{1}{2}} \eta(\tau) d\tau\,,\quad -1/2<H<0\,,
\end{equation}
where 
\begin{equation}
\label{aH}
a_H =\frac{\pi^{1/4} D^{1/2} 2^{H+\frac{1}{2}}}{\sqrt{\Gamma(-H) \Gamma \left(H+1/2\right)}}\,.
\end{equation}
A direct calculation shows that the autocorrelation function $\langle x(t_1) x(t_2)\rangle$ of this centered Gaussian process is identical to $\kappa(t_1,t_2)$ in Eq.~(\ref{kappa0}), so the two processes are indeed identical. Here are some details. The calculation of the autocorrelation function $\langle x(t_1) x(t_2)\rangle$  involves a double integral over $\tau_1$ and $\tau_2$. After getting rid of one integral with the help of the delta-function $\delta(\tau_1-\tau_2)$, we are left with a single integral which can be evaluated analytically:
\begin{equation*}
\int_{-\infty}^{t_{\text{min}}} ds\,(t_1-s)^{H-1/2} (t_2-s)^{H-1/2} =
\frac{2^{-2 H-1} \Gamma (-H) \Gamma \left(H+\frac{1}{2}\right)
   |t_1-t_2|^{2 H}}{\sqrt{\pi }}\,,
\end{equation*}
where $t_{\text{min}}=\text{min}(t_1, t_2)$. The coefficient $a_H^2$, where $a_H$ is defined in Eq.~(\ref{aH}), takes care of the prefactors, leading to the final expression in Eq.~~(\ref{kappa0}).

\subsection{Smoothed fBm for $-1/2<H<0$}
\label{smoothfBm}

The non-point-wise extended fBm can be treated as a Schwartz distribution. Therefore, in the spirit of earlier works  \cite{SmithMeersonSasorov,BM2024b,Monthus2026},  we introduce a \emph{smoothed} extended fBm, $x_{\Delta}(t)$, constructed through local temporal averaging of the bare process $x(t)$ with a narrow filter around each point of time $t$:
\begin{equation}\label{xdelta}
x_{\Delta} (t) \equiv \int\limits_{-\infty}^{\infty} g(t-\tau) x(\tau) d\tau\,.
\end{equation}
Here $g(...)$ is a local filter function normalized to $1$, $\int_{-\infty}^{\infty} g(z) dz =1$, which falls off sufficiently rapidly on a characteristic time scale $\Delta$.
Due to the linearity of the integral operator in  Eq.~(\ref{xdelta}), the smoothed process is Gaussian, and it is fully characterized by its autocorrelation function 
\begin{equation}\label{kappaDelta}
\kappa_{\Delta}(t_1-t_2)=\langle  x_{\Delta} (t_1) x_{\Delta} (t_2)\rangle =\int_{-\infty}^{\infty} d\tau_1 \int _{-\infty}^{\infty} d\tau_2 \,g(t_1-\tau_1) g(t_2-\tau_2) \kappa(\tau_1-\tau_2)\,,
\end{equation}
where $\kappa(t_1-t_2)$ is the autocorrelation function of the ``bare" process presented in Eq.~(\ref{kappa0}).  For concrete calculations we will adopt a Gaussian filter function 
\begin{equation}\label{filter}
g(t) = \frac{e^{-\frac{t^2}{2 \Delta
   ^2}}}{\sqrt{2 \pi \Delta ^2}}\,,
\end{equation}
although the main properties of the smoothed process are insensitive to the filter's details, see below.
To evaluate the double integral 
in Eq.~(\ref{kappaDelta}), we use the spectral density  of the bare process:
\begin{equation}\label{Komega}
K(\omega) = \int_{-\infty}^{\infty} \kappa(\tau) e^{i \omega \tau} d\tau = -2 D \sin (\pi  H) \Gamma (2 H+1) |\omega |^{-2 H-1}\,,
\end{equation}
which has an additional factor $|\omega|^{-2}$ compared with Eq.~(\ref{comegaless}).  Now, replacing $\kappa (\tau_1-\tau_2)$ in Eq.~(\ref{kappaDelta}) by its inverse Fourier transform, $\kappa(\tau) =(2\pi)^{-1} \int_{-\infty}^{\infty} K(\omega) e^{-i\omega \tau}\,d\omega$ and using Eq.~(\ref{filter}),  one can easily evaluate the resulting integrals over $\tau_1$ and $\tau_2$:
$$
\int_{-\infty}^{\infty}  d \tau_1 \int_{-\infty}^{\infty} d\tau_2 \,g(t_1-\tau_1) \, g(t_2-\tau_2) e^{- i\omega (t_1-t_2)} = e^{-\Delta^2 \omega^2-i\omega (t_1-t_2)}\,.
$$
What remains is to evaluate the integral over $\omega$, and we finally obtain
\begin{equation}\label{kappaDelta1}
\kappa_{\Delta}(\tau) = \frac{D\,(2\Delta)^{2 H}  \Gamma\left(H+1/2\right) \,
   _1F_1\left(-H;\frac{1}{2};-\frac{\tau
   ^2}{4 \Delta ^2}\right)}{\sqrt{\pi }}\,,\quad -1/2<H<0\,,
\end{equation}
where $_1F_1\left(a;b;z\right)$ is the Kummer confluent hypergeometric function \cite{NIST}. The spectral density $K_{\Delta}(\omega)$ of the smoothed fBm,
\begin{equation}\label{KDeltaomega}
K_{\Delta}(\omega) =-2 D \sin(\pi  H) \Gamma (2 H+1) e^{-\Delta ^2 \omega ^2} |\omega|^{-2 H-1}\,, \quad -1/2<H<0\,,
\end{equation}
differs from $K(\omega)$ from Eq.~(\ref{Komega}) only by the factor $e^{-\Delta ^2 \omega ^2}$. This factor is crucial at high frequencies, which correspond to short times.

Let us consider some limits of Eq.~(\ref{kappaDelta1}). Setting $\tau =0$, we obtain the variance of the smoothed extended fBm,
\begin{equation}\label{variance}
\text{Var}(H,\Delta) =\frac{D\,(2\Delta)^{2 H} \Gamma \left(H+1/2\right)}{\sqrt{\pi}}\,, \quad -1/2<H<0\,.
\end{equation}
As $\Delta$ goes to zero, the variance diverges. This is of course
to be expected, since the limit of $\Delta\to 0$ brings us back to the bare process. 

The variance (\ref{variance}) also diverges, as $\sim (H+1/2)^{-1}$, at $H\to -1/2$, where the autocorrelation function $\kappa(\tau)$ of the bare process becomes $D/|\tau|$. An inspection of Eq.~(\ref{Komega}) shows that, as $H$ approaches $-1/2$, the spectral density
$K(\omega)$ of the bare fBm becomes $(H+1/2)^{-1} D$, which is independent of $\omega$. Correspondingly, the quantity $(H+1/2)\kappa(\tau)$ behaves as $D \delta(\tau)$.

Importantly, the scaling behavior (\ref{variance}) of the variance with $\Delta$ is quite universal. More precisely, for any normalizable filter function $g(t)$ with a characteristic time scale $\Delta$ the variance of the smoothed fBm behaves as
$\alpha(H) D \Delta^{2H}$, where the $H$-dependent coefficient $\alpha(H)$ diverges as $\sim (H+1/2)^{-1}$ as $H$ approaches $-1/2$.

Further, the $|\tau|\gg \Delta$ asymptotic of $\kappa_{\Delta}(\tau)$,
\begin{equation}\label{largetau}
\kappa_{\Delta}(|\tau|\gg \Delta) \simeq D |\tau|^{2H}\,, \quad -1/2<H<0\,,
\end{equation}
coincides with the autocorrelation function  (\ref{kappa0}) of the bare process: the smoothing is not felt (and not needed) at sufficiently long times.  This result is of course independent of the choice of the filter function $g(t)$.

Still one more instructive limit is $H\to -0$, where 
the autocorrelation function~(\ref{kappaDelta1}) degenerates into
\begin{equation}\label{corrH0}
\kappa_{\Delta}(\tau)=D
\end{equation}
and depends neither on $\tau$, not on $\Delta$. Equation~
(\ref{corrH0}) also holds for any normalizable  filter function $g(t)$, as follows immediately from Eqs.~(\ref{kappa0}) and~(\ref{kappaDelta}).
Interestingly, Eq.~(\ref{corrH0}) matches continuously at $H=0$
to the $H\to +0$ limit of the autocorrelation function~(\ref{kappafBm}) of the traditional fBm process, defined for $0<H<1$.

Since the $H\to -0$ limit~(\ref{corrH0}) of the autocorrelation function is independent of $\tau$, the variance of the smoothed fBm in this limit is also equal to
\begin{equation}\label{varH0}
\text{Var}(0,\Delta)=D\,.
\end{equation}
It is instructive to compare this expression with the $H\to +0$ limit of the variance of a different but related stationary Gaussian process: the fractional
Ornstein-Uhlenbeck (fOU) process \cite{Cheredito2003,Kaarakka2015}, which describes a ``standard" fBm confined by a quadratic external potential. 
The fOU process is defined by the non-Markovian Langevin equation
\begin{equation}\label{fOUeq}
 \dot{x} = -k x+ \xi(t)\,, \quad  0<H<1\,,
\end{equation}
where $\xi(t)$ is the standard fGn, and $k>0$.  The variance of this stationary process is known \cite{Metzler2021,MS2024}:
\begin{equation}\label{varfOU}
\text{Var}_{\text{fOU}}(H,k)= \frac{D \Gamma(2H+1)}{k^{2H}}\,,\quad 0<H<1\,.
\end{equation}
As $H\to +0$, this expression simplifies to $\text{Var}_{\text{fOU}}(0,k)=D$ which coincides with  Eq.~(\ref{varH0}). This remarkable agreement  
occurs because,  at the borderline value $H=0$, both the $\Delta$-dependence in Eq.~(\ref{variance}), and the $k$-dependence in Eq.~(\ref{varfOU}) drop out. 

\begin{figure}[ht]
  \includegraphics[width=0.4\textwidth]{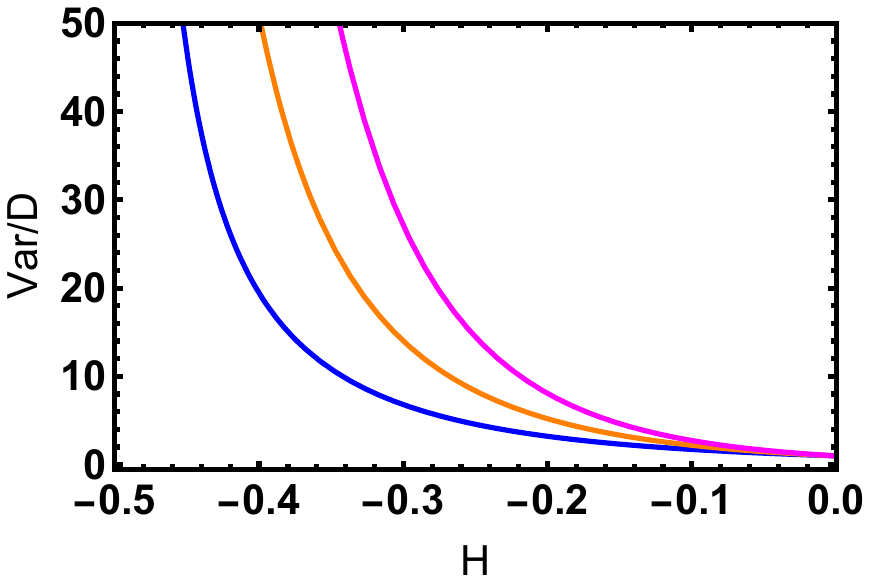}
    \includegraphics[width=0.4\textwidth]{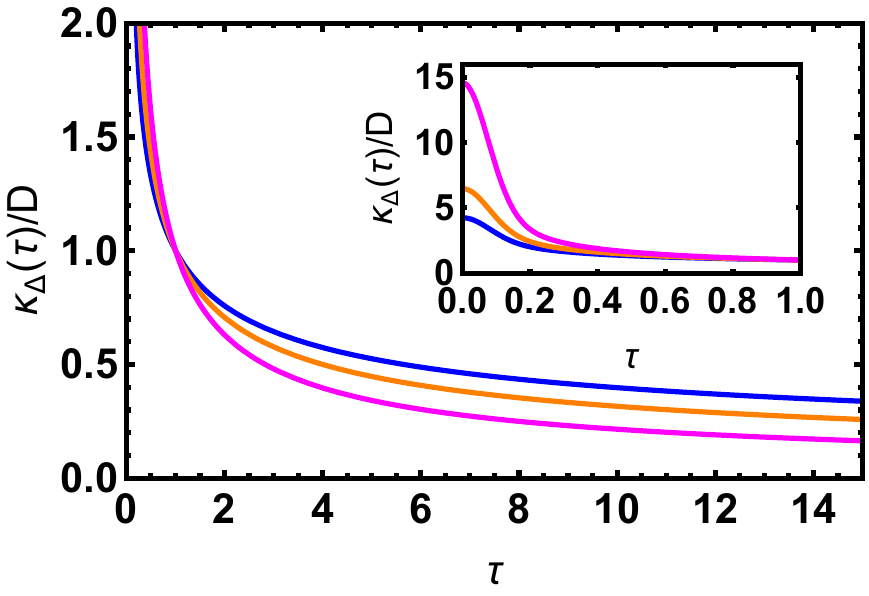}
  \caption{Left: The variance of the smoothed fBm versus $H$, as predicted by Eq.~(\ref{variance}), for $\Delta=0.1$, $0.03$ and $0.01$ (from bottom to top). At $H\to 0$ the three graphs converge to the same point $\text{Var}/D=1$. Right: the autocorrelation function~(\ref{kappaDelta1}) versus $\tau$ for three different values of $H=-1/5,-1/4$ and $-1/3$, and $\Delta=0.05$. The inset emphasizes the small-$\tau$ region.}
  \label{varfig}
\end{figure}

The left panel of Fig. \ref{varfig}  depicts the variance (\ref{variance}) of the smoothed fBm as a function of $H$ for three different values of $\Delta$.  The right panel shows the autocorrelation function~(\ref{kappaDelta1}) versus $\tau$ for $\Delta=0.05$ and three different values of $H$ from the interval $-1/2<H<0$. 

\subsection{Smoothed fGn for $-1/2<H<0$}
\label{smoothfGn}

\begin{figure}[ht]
  \includegraphics[width=0.4\textwidth]{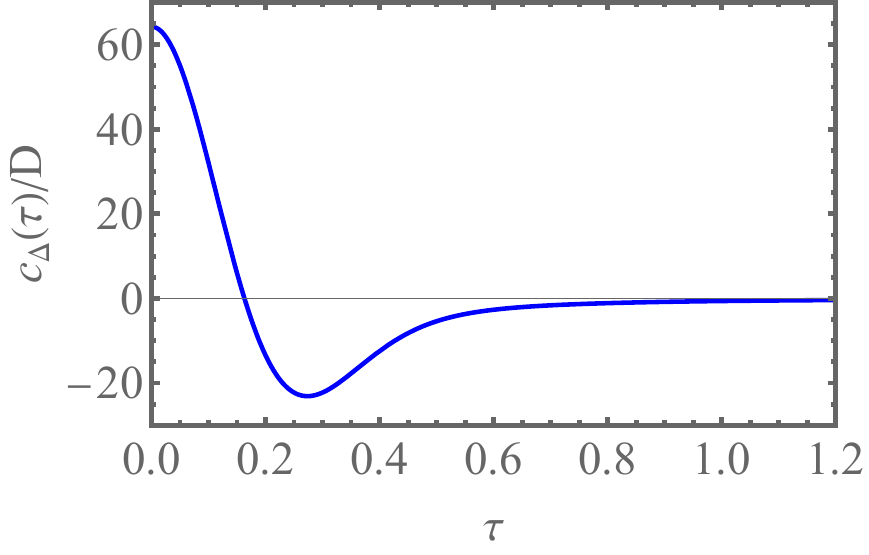}
  \caption{The autocorrelation function $c_{\Delta}(\tau)$ of the smoothed fGn, see Eq.~(\ref{regfGn}), for $H=-1/5$ and $\Delta=1/10$.}
  \label{creg}
\end{figure}

Once the smoothed fBm $x_{\Delta}(t)$ is defined, it is interesting to identify the smoothed fGn $\xi_{\Delta}(t)$, which drives this fBm as described by the Langevin equation 
\begin{equation}\label{Langevin10}
\frac{dx_{\Delta}(t)}{dt} = \xi_{\Delta}(t)\,.
\end{equation}
Since  $\xi_{\Delta}(t)$  is a centered stationary  Gaussian process, it is fully determined by its autocorrelation function $c_{\Delta}(t_1-t_2,H,\Delta)$. For the Gaussian filter (\ref{filter}), $c_{\Delta}(t_1-t_2,H,\Delta)$ can be 
obtained by differentiating the autocorrelation function (\ref{kappaDelta1}) of the smoothed fBm:
\begin{eqnarray}\label{regfGn}
c_{\Delta}(t_1-t_2,H,\Delta) &\equiv& \langle \xi_{\Delta}(t_1) \xi_{\Delta}(t_2) \rangle = \Bigl\langle \frac{d x_{\Delta}(t_1)}{d t_1} \frac{d x_{\Delta}(t_2)}{d t_2}\Bigr\rangle = \frac{\partial^2}{\partial t_1\partial t_2} \kappa_{\Delta}(t_1-t_2) \nonumber \\
&=&\!-\frac{4^H H D \Delta^{2 H-4} \Gamma\left(H+\frac{1}{2}\right)\!\left[(H-1)\tau^2 \,_1F_1\left(2-H;\frac{5}{2};-\frac{\tau^2}{4 \Delta^2}\right)\!+\!3 \Delta ^2 \,_1F_1\left(1-H;\frac{3}{2};-\frac{\tau^2}{4 \Delta^2}\right)\right]}{3 \sqrt{\pi }}\!,
\end{eqnarray}
where $\tau = t_1-t_2$.  Figure \ref{creg} shows a plot of $c_{\Delta}(\tau)$ as a function of $\tau$ for $H=-1/5$ and $\Delta=1/10$. 
Again, the most important properties of $c_{\Delta}(\tau)$, such as the large-$\tau$ asymptotic and the scaling behavior of the variance  with $\Delta$ are insensitive to the particular choice of the filter.

Note that, as $H\to 0$, $c_{\Delta}(\tau)$ from Eq.~(\ref{regfGn}) vanishes identically for all $\tau$, although the autocorrelator of the smoothed extended fBm $\kappa_{\Delta}(\tau)$ does not vanish in this limit, see Eq.~(\ref{corrH0}).  To resolve this apparent contradiction, we note
that $\kappa_{\Delta}(\tau)$ can be written as an improper double integral of $c_{\Delta}(\tau)$. The apparent contradiction is a consequence of non-commutativity of passing to a limit (in this case of $H\to 0$) and integration.

\begin{figure}[ht]
  \includegraphics[width=0.4\textwidth]{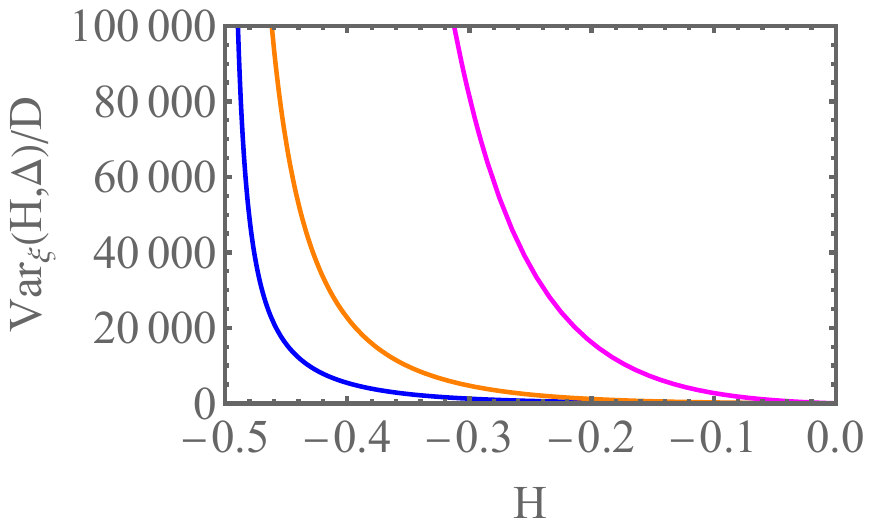}
  \caption{The variance  $\text{Var}_{\xi}(H,\Delta)$ of the smoothed fGn, as described by Eq.~(\ref{varfGn}), as a function of $H$ for $\Delta=0.05$, $0.03$ and $0.01$ (from bottom to top).}
  \label{variancexi}
\end{figure}

The variance of the smoothed fGn $\xi_{\Delta}(t)$, as follows from Eq.~(\ref{regfGn}), is 
\begin{equation}\label{varfGn}
\text{Var}_{\xi}(H,\Delta) \equiv c_{\Delta}(t,t) = \frac{4^H D \Delta ^{2 H-2} |H| \Gamma\left(H+\frac{1}{2}\right)}{\sqrt{\pi}}\,.
\end{equation}
Comparing it with the variance of the smoothed fBm, Eq.~(\ref{variance}), we obtain a simple relation
\begin{equation}\label{Varratio}
\frac{\text{Var}_{\xi}(H,\Delta)}{\text{Var} (H,\Delta)} = \frac{|H|}{\Delta^2} \,.
\end{equation}
Figure \ref{variancexi} depicts $\text{Var}_{\xi}(H,\Delta)$ as a function of $H$ for three values of $\Delta$.  

\begin{figure}[ht]
  \includegraphics[width=0.4\textwidth]{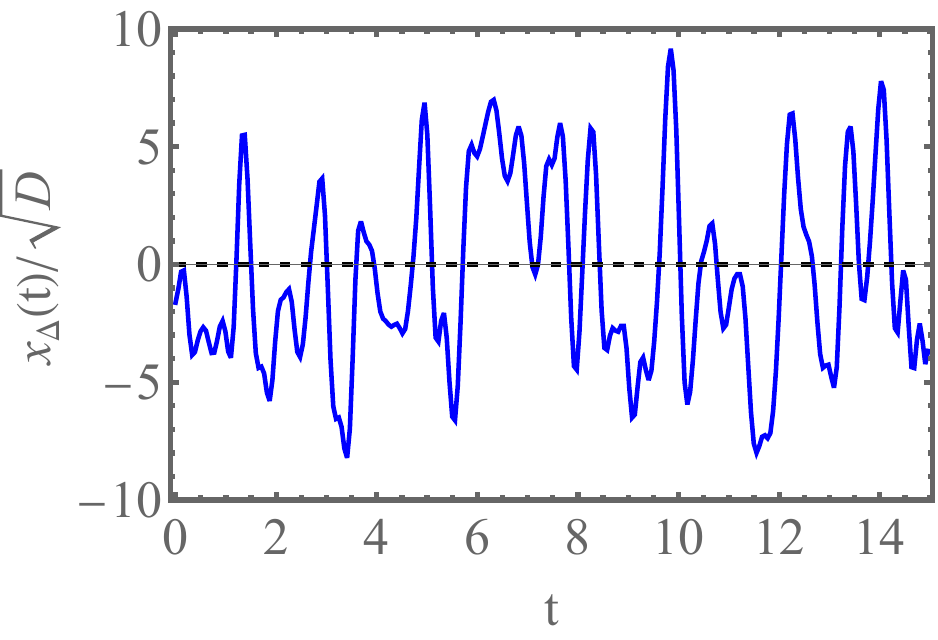}
  \includegraphics[width=0.4\textwidth]{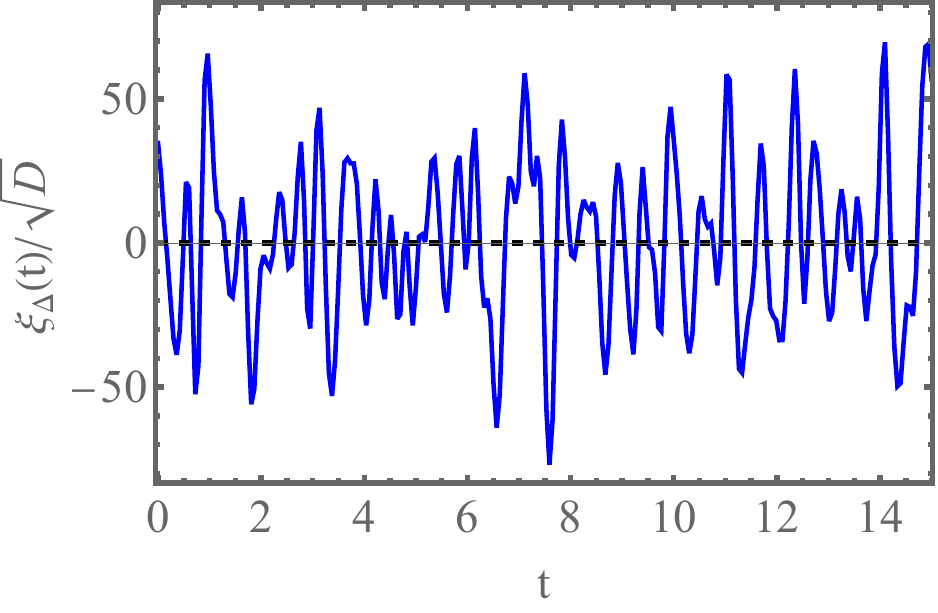}
  \caption{Examples of stochastic realizations of the (time-discretized)  smoothed processes -- the fBm (left panel) and the fGn (right panel) -- for $H=-2/5$ and $\Delta=0.1$.}
  \label{realizations}
\end{figure}

In its turn, Fig.~\ref{realizations} shows examples of stochastic realizations of the (time-discretized) smoothed fBm and fGn processes for $H=-1/5$ and $\Delta=1/10$ as described by the autocorrelation functions (\ref{kappaDelta1}) and (\ref{regfGn}), respectively. These realizations were generated using the discretized covariance method \cite{Rasmussen}.

\section{Optimal fluctuations of the smoothed extended fBm}
\label{OFM}

An additional insight into the  smoothed fBm with negative $H$ is provided by the optimal fluctuation method (OFM), also called  ``geometrical optics" \cite{MS2019,MO2022,HM2024}. As we will see shortly,
the OFM exactly reproduces our result of Secs. \ref{smoothfBm} for the variance of this process.  It also predicts
the optimal -- that is, the most likely -- path of the conditioned smoothed process which dominates the distribution tails of the process. 

A good starting point of the OFM for the smoothed extended fBm is a path integral of the bare extended fBm $x(t)$, which involves the Gaussian action functional \cite{Zinn}
\begin{equation}\label{gaussactiondirect}
\mathcal{S}[x(t)] =\frac{1}{2} \int_{-\infty}^{\infty} dt_1 \int_{-\infty}^{\infty} dt_2\,
  K(t_1-t_2)x(t_1) x(t_2) \,.
\end{equation}
Here $K(t_1-t_2)$ is the inverse kernel to the autocorrelation function $\kappa(t_1-t_2)$ of the bare fBm, presented in Eq.~(\ref{corrless}). The inverse kernel  is defined by the relation
\begin{equation}\label{inversekerneldef}
\int_{-\infty}^{\infty} dt''\,K(t-t'') \kappa (t'-t'') = \delta(t-t')\,.
\end{equation}
As we will see shortly, the explicit form of this kernel $K$ will not be needed, as it also happens for some optimal paths of other Gaussian processes
\cite{Meerson2022,MO2022,LeviVM}.

We condition the \emph{smoothed} process $x_{\Delta}(t)$  on reaching  a specified value $X$. Without losing generality, we assume that the process starts from zero:
\begin{equation}\label{BCs}
x_{\Delta}(t\to -\infty)=0\,.
\end{equation}
This implies that $x(t \to -\infty) = 0$ for the bare process as well.  Since the process is stationary, we can arbitrarily set the time of reaching the value $X$ to $t=0$ and demand 
\begin{equation}\label{X}
x_{\Delta}(t=0) =\int\limits_{-\infty}^{\infty} g(t') x(t') dt' = X.
\end{equation}
The (symmetric) large-$|X|$ tails of the probability density $\mathcal{P}(X)$ are dominated
by the \emph{optimal path} $x(t)$ which minimizes the Gaussian action (\ref{gaussactiondirect})  subject to the condition (\ref{X}) and the condition 

It is convenient to accommodate the condition~(\ref{X}) by introducing a modified action 
\begin{equation}\label{modified}
\mathcal{S}_{\lambda}[x(t)] = \mathcal{S}[x(t)]-\lambda \int_{-\infty}^{\infty}g(t) x(t) \, dt\,,
\end{equation}
where $\mathcal{S}[x(t)]$ is given by Eq.~(\ref{gaussactiondirect}). The Lagrange multiplier $\lambda$ is to be ultimately expressed through $X$ and the rest of the parameters of the problem.  Having found the optimal path, we can evaluate the probability distribution $\mathcal{P}(X)\sim \exp(-S)$ up to a pre-exponential factor by determining the action along the optimal path. It is more convenient, however, to use a shortcut in the form of the relation
\begin{equation}\label{shortcut}
 \frac{dS}{dX}=\lambda(X)\,,
\end{equation}
which reflects the fact that $X$ and $\lambda$ are conjugate variables, see, e.g. Ref. \cite{Cunden}. Equation~(\ref{shortcut}) makes it possible to calculate the action once the Lagrange multiplier $\lambda$ is expressed through $X$.

Demanding that the linear variation of the modified action (\ref{modified}) vanish, we arrive at a linear integral equation for the optimal path of the \emph{bare} process:
\begin{equation}\label{ELdirect}
\int_{-\infty}^{\infty} K(t'-\tau) x(\tau) d\tau=\lambda g(t')\,.
\end{equation}
This equation can be easily solved by multiplying the both sides  by $\kappa(t-t')$, integrating over $t'$ from $-\infty$ to $\infty$, and using the definition (\ref{inversekerneldef}) of the inverse kernel. The resulting solution for the optimal path (which we also denote by $x(t)$) is
\begin{equation}\label{xlambdadirect}
x(t) = \lambda \int_{-\infty}^{\infty} g(t') \kappa(t-t') dt' = \frac{2^H D \lambda  \Delta ^{2 H} \Gamma \left(H+\frac{1}{2}\right) \,
   _1F_1\left(-H;\frac{1}{2};-\frac{t^2}{2 \Delta ^2}\right)}{\sqrt{\pi }}\,,\quad -1/2<H<0\,.
\end{equation}
As one can check,  the boundary condition (\ref{BCs}) is obeyed automatically. Note that the first equality in Eq.~(\ref{xlambdadirect}) holds for any autocorrelation function $\kappa(\tau)$ and any filter $g(t)$.

\begin{figure}[ht]
  \includegraphics[width=0.4\textwidth]{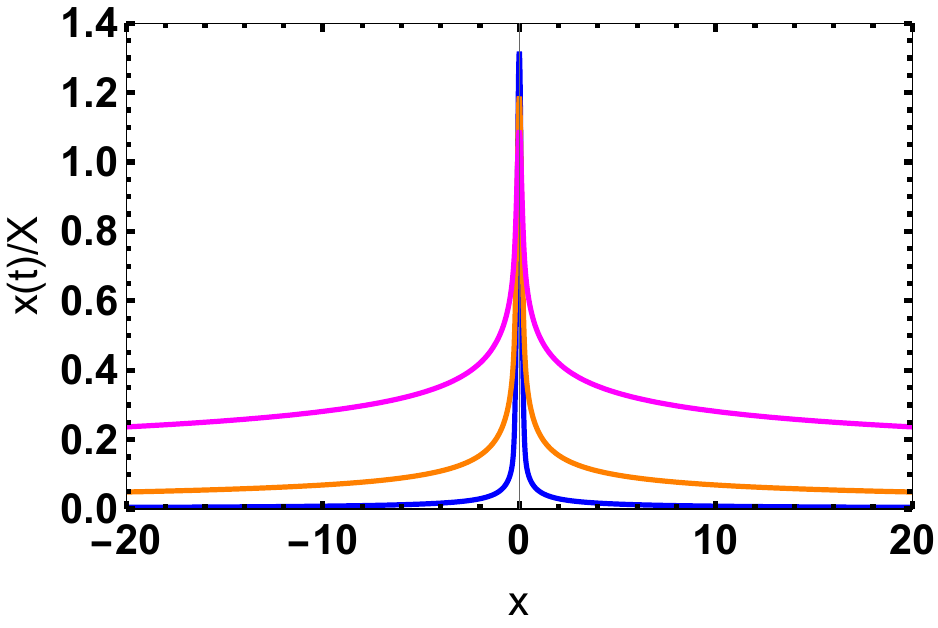}
  \includegraphics[width=0.4\textwidth]{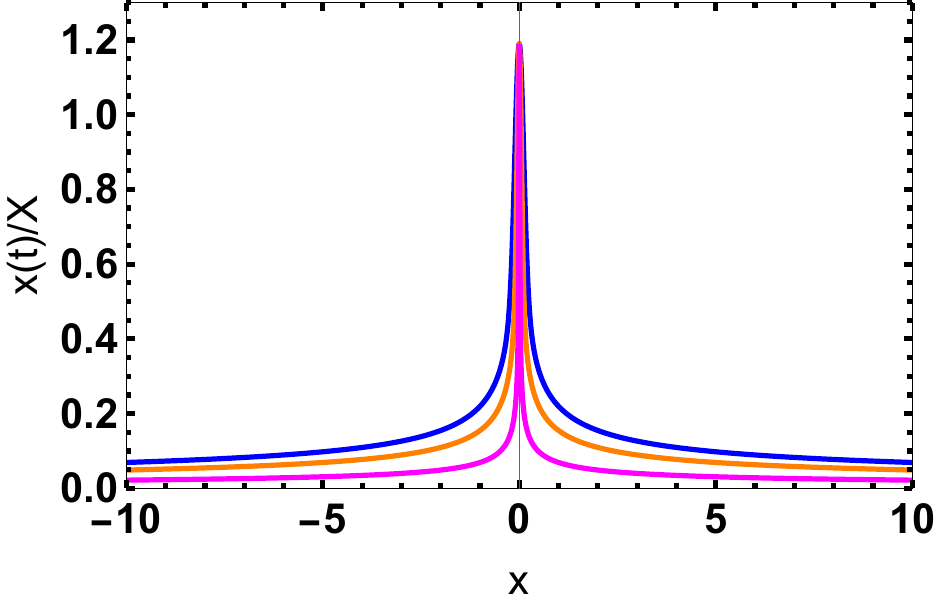}
  \caption{The optimal path $x(t)$ as predicted by Eq.~(\ref{xopt}). Left panel: at fixed $\Delta=0.1$ and three values of $H$: $-2/5$, $-1/4$ and $-1/8$ (from bottom to top).
  Right panel: at fixed  $H=-1/4$ and three values of $\Delta$: $0.1$, $0.05$ and $0.01$ (from top to bottom).}
  \label{figa}
\end{figure}

Now we evaluate 
\begin{equation}\label{intX}
\int\limits_{-\infty}^{\infty} g(t) x(t) dt = -\frac{D \lambda  \Delta ^{2 H} \sin (\pi  H) \Gamma (-H) \Gamma (2 H+1)}{\pi } 
\end{equation}
and determine $\lambda$ from Eq.~(\ref{X}). We obtain
\begin{equation}\label{lambdavsX} 
 \lambda=\lambda(X)=\frac{\sqrt{\pi}(2\Delta)^{-2 H} X }{D \Gamma
   \left(H+1/2\right)}\,,\quad -1/2<H<0\,.
\end{equation}
As a result, the  optimal path of the bare process is the following:
\begin{equation}\label{xopt}
\frac{x(t)}{X} = 2^{-H} \, _1F_1\left(-H;\frac{1}{2};-\frac{t^2}{2 \Delta ^2}\right)\,,\quad -1/2<H<0\,.
\end{equation}
In particular, its short- and long-time asymptotics  are 
\begin{equation}\label{asympx}
  \frac{x(t)}{X} =
    \begin{cases}
         2^{-H}\left(1+\frac{ H t^2}{\Delta ^2}+\dots\right)\,, & t\to 0\,,\\
   \frac{2^{-2H} \sqrt{\pi }}{\Gamma \left(H+1/2\right)} \left(\frac{|t|}{\Delta }\right)^{2
   H}+\dots\,, & |t|\to \infty\,.
    \end{cases}
\end{equation}
Note that, although the bare process itself is not a point-wise function  (recall that we consider $-1/2<H<0$), the optimal path of the bare process is not only a point-wise function, but it is also regular.  As one can see from the second line of Eq.~(\ref{asympx}), $x(t)$ goes to zero at $t\to \pm \infty$, which verifies a posteriori the boundary condition~(\ref{BCs}). Further,  the bare optimal path reaches its maximum at $t=0$, and the value of the maximum is independent of the regularization time scale $\Delta$.  The left panel of Fig. \ref{figa} shows the bare optimal path $x(t)$ at fixed $\Delta=0.1$ and three different values of $-1/2<H<0$. The right panel of the same figure shows
the bare optimal path at fixed $H=-1/4$ and three different values of $\Delta$.

Now we can calculate the action by using Eqs.~(\ref{shortcut}) and (\ref{lambdavsX}):
\begin{equation}\label{actionm=2}
S(X,H,\Delta)=\int_0^{X} dX\, \lambda(X) = \frac{2^{-2 H-1} \sqrt{\pi } \Delta ^{-2 H}  X^2}{D \Gamma
   \left(H+1/2\right)}\,,\quad -1/2<H<0\,.
\end{equation}
Since the process is Gaussian, what is left is to normalize the distribution $\mathcal{P}(X) \sim \exp[-S(X,H,\Delta)]$ to unity. As to be expected, the variance of $X$, as described by Eq.~(\ref{actionm=2}), perfectly agrees with the exact result in Eq.~(\ref{variance}).

It is also of interest to determine the optimal path $x_{\Delta}(t)$ of the smoothed process defined by Eq.~(\ref{xdelta}). A straightforward way to do it is to plug Eq.~(\ref{xopt}) into Eq.~(\ref{xdelta}) and evaluate the resulting integral by using the direct and inverse Fourier transforms. A more elegant way is to use the fact that,  in terms of the optimal path of the smoothed process,  the Gaussian action has the form of Eq.~(\ref{gaussactiondirect}) with $K(t_1-t_2)$ replaced by  $K_{\Delta}(t_1-t_2)$: the inverse kernel for the autocorrelation function~$\kappa_{\Delta}(t_1-t_2)$  given by Eq.~(\ref{kappaDelta1}). Rewriting the condition $x_{\Delta}(t=0) = X$ as an integral constraint
\begin{equation}\label{regcondition}
\int_{-\infty}^{\infty} x_{\Delta}(t) \delta(t) dt = X\,,
\end{equation}
and performing the minimization of the modified action, we arrive at the integral equation
\begin{equation}\label{ELdirect2}
\int_{-\infty}^{\infty} K_{\Delta}(t-t') x_{\Delta}(t') dt' =\lambda \delta(t)\,.
\end{equation}
Its solution is obvious:
\begin{equation}\label{xlambdadirect1}
x_{\Delta}(t) = \lambda \kappa_{\Delta}(t)\,,
\end{equation}
and it holds for any reasonable filter. Now, using Eq.~(\ref{kappaDelta1}) and determining $\lambda$ from the condition $x_{\Delta}(t=0) = X$, we obtain the final result for the Gaussian filter:
\begin{equation}\label{xdeltaopt}
  \frac{x_{\Delta}(t)}{X}= \, _1F_1\left(-H;\frac{1}{2};-\frac{t^2}{4 \Delta ^2}\right)\,.
\end{equation}
Comparing the optimal path (\ref{xdeltaopt}) of the smoothed process with the optimal path of the bare process, Eq.~(\ref{xopt}), we notice a difference in the third arguments of the Kummer function, and the presence of the additional factor $2^{-H}$ in  Eq.~(\ref{xopt}). 
Notably, the normalized ratio $x_{\Delta}(t)/x_{\Delta}(0)$, as described by Eq.~(\ref{xdeltaopt}), perfectly coincides with the normalized autocorrelation function $\kappa_{\Delta}(t)/\kappa_{\Delta}(0)$, see Eq.~(\ref{kappaDelta1}). This is just one more instance of
a general property of optimal paths of Gaussian processes, conditioned on reaching a specified value at a point \cite{Meerson2022,MO2022,MS2024,LeviVM}.

\section{Fractional Ornstein-Uhlenbeck process for $-1/2<H<0$}
\label{fOU}

One can extend the standard fOU process, defined by Eq.~(\ref{fOUeq}),
to the region of $-1/2<H<0$ in a similar way, by using the extended fGn $\xi(t)$. The autocorrelation function of the bare fOU process can be obtained
by a straightforward calculation in the Fourier space. Starting from Eq.~(\ref{fOUeq}), one obtains
\begin{equation}\label{kappafOUbare0}
\kappa^{\text{fOU}}(\tau)=\frac{1}{2\pi} \int_{-\infty}^{\infty} d\omega \,c(\omega)\int_{-\infty}^{t_1} d\tau_1\int_{-\infty}^{t_2} d\tau_2
\exp\left[-k(t_1-\tau_1+t_2-\tau_2) - i \omega (\tau_1-\tau_2)\right]\,,
\end{equation}
where $c(\omega)$ -- the spectral density of the fGn -- is given by Eq.~(\ref{comegaless}). The integrals over $\tau_1$ and $\tau_2$ are elementary. Evaluating the remaining integral over $\omega$, we arrive at
\begin{equation}\label{kappafOUbare}
\kappa^{\text{fOU}}(\tau) =D \left[|\tau|^{2H} \,_1F_2\left(1;H+\frac{1}{2},H+1;\frac{k^2 \tau ^2}{4}\right)-k^{-2 H}\Gamma (2 H+1) \cosh (k \tau)\right]\,,\quad -1/2<H<0\,.
\end{equation}
This expression also follows from a formal extension to the range $-1/2<H<0$ of the autocorrelation function for the standard fOU process, $0<H<1$, obtained in Refs. \cite{Cheredito2003,MS2024}. One only needs to change the sign there so that the spectral density of the fGn remains positive, as in Eq.~(\ref{corrless}).

The variance $\kappa^{\text{fOU}}(\tau=0)$ of this stationary Gaussian process  diverges, hence we define a smoothed process $x^{\text{fOU}}_{\Delta}(t)$ by using  Eq.~(\ref{xdelta}): for concreteness, with the same Gaussian filter function~(\ref{filter}) as before. Calculating the variance of the smoothed process, we obtain
\begin{equation}\label{vardeltafOU}
 \text{Var}^{\text{fOU}}(H,k,\Delta) =  \frac{4^H |H| D k^{-2 H} \Gamma\left(H+1/2\right) e^{ k^2\Delta^2} \Gamma \left(H,k^2 \Delta^2\right)}{\sqrt{\pi }} \,,
 \quad -1/2<H<0\,.
\end{equation}
where $\Gamma(a,z)$ is the incomplete gamma function.  As to be expected, for $k=0$ Eq.~(\ref{vardeltafOU}) brings us back to Eq.~(\ref{variance}) for the smoothed fBm.
The variance (\ref{vardeltafOU}) diverges at $\Delta \to 0$, also as to be expected for a non-point-wise bare process. 

Of most interest, however, is the asymptotic of $0<k \Delta \ll 1$, when the regularization time scale $\Delta$ is nonzero but very small compared with the characteristic relaxation  time $1/k$ of the fOU process. In the leading order, this asymptotic,
\begin{equation}\label{smallDeltafOU}
 \text{Var}^{\text{fOU}}(H,k,\Delta\ll 1/k)\simeq \frac{D (2\Delta)^{2 H} \Gamma\left(H+1/2\right)}{\sqrt{\pi}}\,, \quad -1/2<H<0\,,
\end{equation}
is independent of $k$ and perfectly coincides with the small-$\Delta$ variance for the pure smoothed fBm, see Eq.~(\ref{variance}). This asymptotic insensitivity of the smoothed extended fOU process to the strength of the confinement potential is both unexpected and remarkable.

The variance (\ref{smallDeltafOU}) also matches continuously with the variance of the standard fOU process, $0<H<1$, at the borderline value $H=0$\footnote{\label{footnoteH0}However, if one takes the limit of $H\to 0$  at fixed $k\Delta>0$ in Eq.~(\ref{vardeltafOU}), the resulting variance vanishes.}. Finally, the variance of the smoothed fOU process diverges at $H=-1/2$.

\begin{figure}[ht]
  \includegraphics[width=0.4\textwidth]{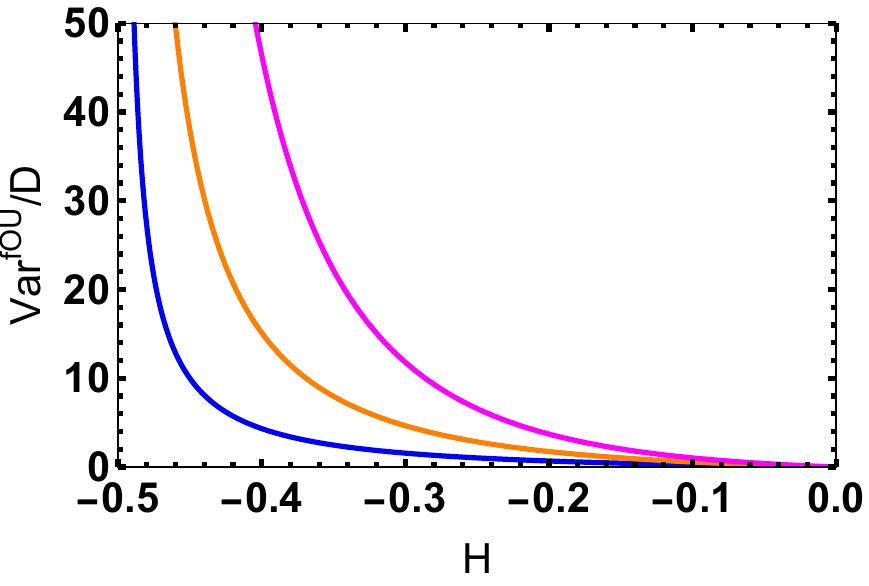}
  \caption{The variance of the smoothed fOU process versus $H$, as predicted by Eq.~(\ref{vardeltafOU}), for $k=1$ and $\Delta=0.3$, $0.1$ and $0.03$, from bottom to top.}
  \label{variancefOU}
\end{figure}

Figure \ref{variancefOU} shows the $H$-dependence of the variance~(\ref{vardeltafOU}) of the smoothed fOU process for $k=1$ and three different values of $\Delta$.

We also calculated, for the Gaussian filter, the spectral density (the  Fourier transform of the autocorrelation function) of the smoothed fOU process:
\begin{equation}\label{FTKfOU}
\tilde{\kappa}^{\text{fOU}}_{\omega} =\frac{2 D e^{-\Delta ^2 \omega ^2} \sin (\pi|H|) \,\Gamma (2 H+1) |\omega | ^{1-2 H}}{k^2+\omega ^2}\,.
\end{equation}
Again, the regularization is present here only through the factor $e^{-\Delta ^2 \omega ^2}$ which is important at high frequencies.
The autocorrelation function $\kappa^{\text{fOU}}(\tau)$ itself is given by the inverse Fourier transform,
$\tilde{\kappa}^{\text{fOU}}(\tau) = (2\pi)^{-1} \int_{-\infty}^{\infty} \tilde{\kappa}^{\text{fOU}}_{\omega} e^{i\omega \tau} d\omega$\,.

\section{Discussion}
\label{discussion}

As we have seen, the fBm extended to the regime $-1/2<H<0$ and smoothed at small scales differs fundamentally from the standard fBm, and these differences lead to distinctive dynamical consequences. In the standard fBm the persistence and the roughness of paths are mutually exclusive. In contrast, the extended fBm is both persistent and rough. It is also stationary, therefore diffusion is completely suppressed. In its turn, the smoothed extended fractional Ornstein-Uhlenbeck (fOU) process exhibits a remarkable and counter-intuitive effect of asymptotic insensitivity to the confinement potential's strength. For the reader's convenience, we summarize the bare and smoothed extended processes, studied in this work, in a table.

\begin{center}
\begin{tabular}{ |c|c|c| } 
 \hline
$k=0,\,\Delta=0$ & bare extended fGn& bare extended fBm\\ 
$k=0,\,\Delta>0$ & smoothed extended fGn& smoothed extended fBm\\ 
\hline
$k>0,\,\,\Delta=0$ & bare extended fOU\\ 
$k>0,\,\,\Delta>0$ & smoothed extended fOU\\ 
 \cline{1-2}
\end{tabular}
\end{center}

Although our regularization calculations employed a particular,  Gaussian filter (\ref{filter}), the most important results -- such as the scaling dependence of the variance of the smoothed extended fBm or fOU process on $\Delta$ -- are insensitive
to the specific choice of the filter.

The Gaussian character of the smoothed extended fBm makes this process realizable by standard techniques. This brings us to
a challenging yet promising direction for future work:  an experimental test of our prediction of complete diffusion suppression in the range 
$-1/2<H<0$. Such an experiment would require tracking the motion of overdamped particles driven by smoothed fractional Gaussian noise (fGn) in the range $-1/2<H<0$, with the crucial condition that the externally applied fGn dominates over thermal noise. A natural experimental platform for such a study is optically driven colloidal suspensions \cite{Roichman}\footnote{\label{36beams}{In Ref. \cite{Roichman} a nearly spatially uniform optical driving of colloidal particles was achieved by using 36 laser beams.}}. The optical driving can be controlled by computer-generated stochastic realizations of smoothed fGn, such as those shown in Fig.~\ref{realizations}.

It can be useful to extend the definition of the smoothed fBm for $-1/2<H<0$ to include the point $H=0$ as well. For the traditional fBm in the range $0<H<1$ a robust extension to $H=0$ was done (via a different regularization) in Refs. \cite{Fyodorov,Chevillard,NeumanRosenbaum}, leading to log-correlated processes.

In conclusion, the smoothed extended fBm -- an almost scale-invariant stationary Gaussian process which is both persistent and rough -- provides us with a convenient and versatile mathematical model and will hopefully find applications in a simplified modeling of real-world processes and fields -- in physics, Earth sciences, climate studies, biology,  finance and other areas.

\section*{Acknowledgments}

We are grateful to Yael Roichman and Naftali R. Smith for useful discussions. B.M. was supported by the Israel Science Foundation (Grant No. 1499/20).


\vspace{0.5 cm}

\begin{thebibliography} {99}


\bibitem{Kolmogorov} A. N. Kolmogorov, C. R. Dokl. Acad. Sci. URSS \textbf{26}, 115
(1940).

\bibitem{Mandelbrot} B. B. Mandelbrot and J. W. van Ness, SIAM Review \textbf{10}, 422 (1968).

\bibitem{Qian2003} H. Qian, in \textit{``Processes with Long-Range
Correlations. Theory and Applications"}, edited by G. Rangarajan and M. Ding (Springer, Berlin, 2003), p. 22.

\bibitem{Chow} W. C. Chow, WIREs Comp. Stat. \textbf{3}, 149 (2011).

\bibitem{Metzlerreview}  R. Metzler, J.-H. Jeon, A. G. Cherstvy, and E. Barkai, Phys.
Chem. Chem. Phys. \textbf{16}, 24128 (2014).

\bibitem{Monthus2026} C. Monthus, 
J. Stat. Mech. (2026) 023207.

\bibitem{Krug} J. Krug, Adv. Phys. \textbf{46}, 139 (1997).

\bibitem{SmithMeersonSasorov} N. R. Smith,  B.  Meerson and P. V. Sasorov, Phys. Rev. E \textbf{95}, 012134 (2017).

\bibitem{HalpinHealy2014} T. Halpin–Healy and G. Palasantzas, Europhys. Lett. \textbf{105}, 50001 (2014).

\bibitem{Almeida} R. A. L. Almeida, S. O. Ferreira, T. J. Oliveira, and F. D. A. Aar\~{a}o Reis, Phys. Rev. B \textbf{89}, 045309 (2014).

\bibitem{Reis} F. D. A. Aar\~{a}o Reis, J. Stat. Mech. (2015) P11020.

\bibitem{Carrasco} I. S. S. Carrasco and T. J. Oliveira, Phys. Rev. E \textbf{93}, 012801 (2016).

\bibitem{BM2024b} E. Bettelheim and B. Meerson, J. Stat. Mech. (2024) 113204.

\bibitem{NIST} F. W. J. Olver, D. W. Lozier, R. F. Boisvert, and C. W. Clark,
\textit{NIST Handbook of Mathematical Functions} (Cambridge University Press, Cambridge, UK, 2010).

\bibitem{Cheredito2003} P. Cheredito, H. Kawaguchi, and M. Maejima, Electron. J. Probab. \textbf{8}, 1 (2003).

\bibitem{Kaarakka2015} T. Kaarakka, \textit{Fractional Ornstein-Uhlenbeck Processes} (Tampere Univ. Technol., Tampere, Finland, 2015).

\bibitem{Metzler2021} T. Guggenberger, A. Chechkin,  and R. Metzler, J. Phys. A: Math. Theor. \textbf{54}, 29 (2021).

\bibitem{MS2024} B. Meerson and P. V. Sasorov, 
J. Phys. A: Math. Theor. \textbf{57}, 445002 (2024).


\bibitem{Rasmussen} C. E. Rasmussen and C. K. I. Williams, \textit{Gaussian Processes for Machine Learning} (The MIT Press, Boston, 2006).

\bibitem{MS2019} B. Meerson and N. R Smith, J. Phys. A: Math. Theor. \textbf{52}, 415001 (2019).

\bibitem{MO2022} B. Meerson and G. Oshanin,
Phys. Rev. E \textbf{105}, 064137 (2022).

\bibitem{HM2024} A. K. Hartmann and B. Meerson,  
Phys. Rev. E \textbf{109}, 014146 (2024).

\bibitem{Zinn} J. Zinn-Justin, \textit{Quantum Field Theory and Critical Phenomena}, International Series of Monographs on Physics, 4th ed.
(Clarendon, Oxford, UK, 2002).

\bibitem{Meerson2022} B. Meerson, Phys. Rev. E \textbf{105}, 034106 (2022); \textbf{107}, 039902(E) (2023).  


\bibitem{LeviVM} A. Valov, N. Levi, and B. Meerson, Phys. Rev. E \textbf{110}, 024138 (2024).  

\bibitem{Cunden} F. D. Cunden, P. Facchi, and P. Vivo, J. Phys. A: Math.
Theor. \textbf{49}, 135202 (2016).

\bibitem{Roichman} G. Geva, T. Admon, M. Levin, and Y. Roichman, Phys. Rev. Lett. \textbf{134}, 218201 (2025).

\bibitem{Fyodorov} Y. V. Fyodorov, B. A. Khoruzhenko, and N. J. Simm,  Ann. Prob. \textbf{44},  2980 (2016).

\bibitem{Chevillard} L. Chevillard, Phys. Rev. E \textbf{96}, 033111 (2017).

\bibitem{NeumanRosenbaum} E. Neuman and M. Rosenbaum, Electron. Commun. Probab. \textbf{23}, 1-12  (2018).




\end{thebibliography}
\end{document}